%% file: main.tex
\documentclass[manuscript, nonacm]{acmart}

\AtBeginDocument{%
  }

\begin{document}

\title{OriStitch: A Machine Embroidery Workflow to Turn Existing Fabrics into Self-Folding 3D Textiles}

\author{Zekun Chang}
\orcid{0000-0002-4388-3451}
\affiliation{
  \institution{Cornell Tech}
  \city{New York}
  \country{USA}}
\email{zc247@cornell.edu}

\author{Yixuan Gao}
\orcid{0000-0003-1778-3104}
\affiliation{
  \institution{Cornell Tech}
  \city{New York}
  \country{USA}}
\email{yg478@cornell.edu}

\author{Yuta Noma}
\orcid{0000-0001-9654-2573}
\affiliation{%
  \institution{The University of Toronto}
  \city{Toronto}
  \country{Canada}
}
\email{yutanoma@dgp.toronto.edu}

\author{Shuo Feng}
\orcid{0000-0003-2350-321X}
\affiliation{%
  \institution{Cornell Tech}
  \city{New York}
  \country{USA}
}
\email{sf522@cornell.edu}

\author{Xinyi Yang}
\orcid{0000-0002-8877-0813}
\affiliation{%
  \institution{Harvard University}
  \city{Cambridge}
  \country{USA}
}
\email{xinyiyang@gsd.harvard.edu}

% \author{Kazuhiro Shinoda}
% \orcid{0000-0002-2789-0109}
% \affiliation{%
%   \institution{The University of Tokyo}
%   \city{Tokyo}
%   \country{Japan}
% }
% \email{kazuhiro@iis-lab.org}

% \author{Tung D. Ta}
% \orcid{0000-0002-2342-1364}
% \affiliation{%
%   \institution{The University of Tokyo}
%   \city{Tokyo}
%   \country{Japan}
% }
% \email{tung@csg.ci.i.u-tokyo.ac.jp}

\author{Kazuhiro Shinoda}
\email{kazuhiro@iis-lab.org}
\orcid{0000-0002-2789-0109}
\author{Tung D. Ta}
\email{tung@csg.ci.i.u-tokyo.ac.jp}
\orcid{0000-0002-2342-1364}
\affiliation{%
  \institution{The University of Tokyo}
%  \streetaddress{7-3-1 Hongo, Bunkyo-ku}
  \city{Tokyo}
  \country{Japan}
%  \postcode{113-8656}
}

% \author{Koji Yatani}
% \orcid{0000-0003-4192-0420}
% \affiliation{%
%   \institution{The University of Tokyo}
%   \city{Tokyo}
%   \country{Japan}
% }
% \email{koji@iis-lab.org}

% \author{Tomoyuki Yokota}
% \orcid{0000-0003-1546-8864}
% \affiliation{%
%   \institution{The University of Tokyo}
%   \city{Tokyo}
%   \country{Japan}
% }
% \email{yokota@ntech.t.u-tokyo.ac.jp}

% \author{Takao Someya}
% \orcid{000-0002-9458-9033}
% \affiliation{%
%   \institution{The University of Tokyo}
%   \city{Tokyo}
%   \country{Japan}
% }
% \email{someya@ee.t.u-tokyo.ac.jp}

\author{Koji Yatani}
\email{koji@iis-lab.org}
\orcid{0000-0003-4192-0420}
\author{Tomoyuki Yokota}
\email{yokota@ntech.t.u-tokyo.ac.jp}
\orcid{0000-0003-1546-8864}
\author{Takao Someya}
\email{someya@ee.u-tokyo.ac.jp}
\orcid{0000-0003-3051-1138}
\affiliation{%
  \institution{The University of Tokyo}
%  \streetaddress{7-3-1 Hongo, Bunkyo-ku}
  \city{Tokyo}
  \country{Japan}
%  \postcode{113-8656}
}

% \author{Yoshihiro Kawahara}
% \orcid{0000-0002-0310-2577}
% \affiliation{%
%   \institution{The University of Tokyo}
%   \city{Tokyo}
%   \country{Japan}
% }
% \email{kawahara@akg.t.u-tokyo.ac.jp}

% \author{Koya Narumi}
% \orcid{0000-0001-5830-3942}
% \affiliation{%
%   \institution{The University of Tokyo}
%   \city{Tokyo}
%   \country{Japan}
% }
% \email{narumi@akg.t.u-tokyo.ac.jp}

\author{Yoshihiro Kawahara}
\email{kawahara@akg.t.u-tokyo.ac.jp}
\orcid{0000-0003-2042-6597}
\author{Koya Narumi}
\email{narumi@akg.t.u-tokyo.ac.jp}
\orcid{0000-0001-5830-3942}
\affiliation{%
  \institution{The University of Tokyo}
%  \streetaddress{7-3-1 Hongo, Bunkyo-ku}
  \city{Tokyo}
  \country{Japan}
%  \postcode{113-8656}
}

\author{Fran\c{c}ois Guimbreti\`ere}
\orcid{0000-0002-5510-6799}
\authornote{Both authors contributed equally to the paper.}
\affiliation{
  \institution{Cornell University}
  \city{New York}
  \country{USA}}
\email{fvg3@cornell.edu}

\author{Thijs Roumen}
\orcid{0000-0003-2042-6597}
\authornotemark[1]
\affiliation{
  \institution{Cornell Tech}
  \city{New York}
  \country{USA}}
\email{thijs.roumen@cornell.edu}

\renewcommand{\shortauthors}{Chang et al.}

\begin{abstract}

OriStitch is a computational fabrication workflow to turn existing flat fabrics into self-folding 3D structures. It uses machine embroidering of functional threads in specific patterns on fabrics, users then apply heat to deform the structure into a target 3D structure. OriStitch is compatible with a range of existing materials (e.g., leather, woven fabric, and denim). 

We present the design of specific embroidered hinges that fully close under exposure to heat. We discuss the stitch pattern design, thread and fabric selection, and heating conditions. To allow users to create 3D textiles using our hinges, we create a tool to convert 3D meshes to 2D stitch patterns automatically, as well as an end-to-end fabrication and actuation workflow. To validate this workflow, we designed and fabricated a range of application models consisting of up to 338 hinges each (the cap of Figure 1, a handbag, and a vase cover).

In technical evaluation, we found that our tool successfully 26/28 models found in related papers. We also demonstrate the folding performance across different materials (suede leather, cork, Neoprene, and felt). 
\end{abstract}




%%\begin{CCSXML}
%%<ccs2012>
%% <concept>
%%  <concept_id>00000000.0000000.0000000</concept_id>
%%  <concept_desc>Do Not Use This Code, Generate the Correct Terms for Your Paper</concept_desc>
%%  <concept_significance>500</concept_significance>
%% </concept>
%% <concept>
%%  <concept_id>00000000.00000000.00000000</concept_id>
%%  <concept_desc>Do Not Use This Code, Generate the Correct Terms for Your Paper</concept_desc>
%%  <concept_significance>300</concept_significance>
%% </concept>
%% <concept>
%%  <concept_id>00000000.00000000.00000000</concept_id>
%%  <concept_desc>Do Not Use This Code, Generate the Correct Terms for Your Paper</concept_desc>
%%  <concept_significance>100</concept_significance>
%% </concept>
%% <concept>
%%  <concept_id>00000000.00000000.00000000</concept_id>
%%  <concept_desc>Do Not Use This Code, Generate the Correct Terms for Your Paper</concept_desc>
%%  <concept_significance>100</concept_significance>
%% </concept>
%%</ccs2012>
%%\end{CCSXML}

%%\ccsdesc[500]{Do Not Use This Code~Generate the Correct Terms for Your Paper}
%%\ccsdesc[300]{Do Not Use This Code~Generate the Correct Terms for Your Paper}
%%\ccsdesc{Do Not Use This Code~Generate the Correct Terms for Your Paper}
%%\ccsdesc[100]{Do Not Use This Code~Generate the Correct Terms for Your Paper}

\begin{teaserfigure}
  \includegraphics[width=\textwidth]{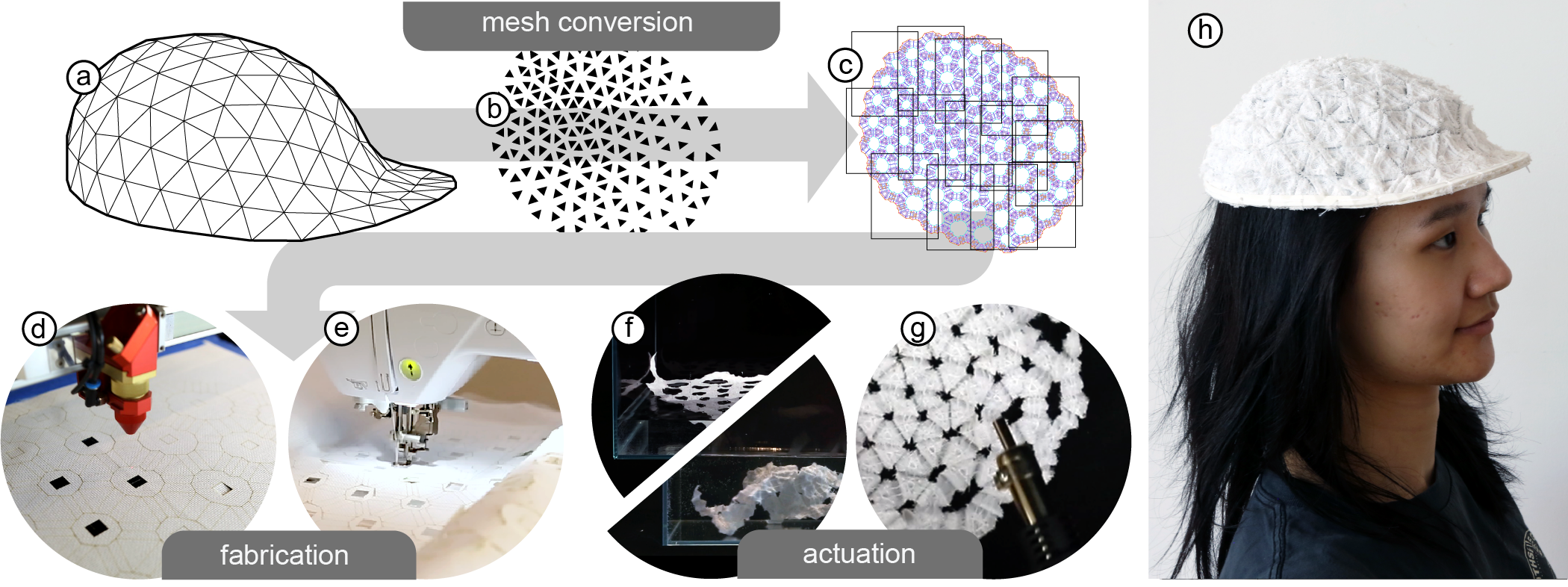}
  \caption{OriStitch turns existing fabrics into self-folding structures using machine embroidery. It consists of a software tool and a fabrication technique. The software tool extends \textit{Origamizer} \cite{Tachi2010:Origamizer} to convert (a)~3D geometry to (b)~2D fold-able faces, (c)~OriStitch places embroidered hinges between faces based on four hinge types. The exported embroidery and laser cutting patterns are used to (d)~score the hinges, remove excess material, and (e)~stitch the hinges onto the fabric. (f)~Users dissolve water soluble support stitches, and start actuation of the heat shrinking threads by pre-creasing the hinges in a hot water bath (g)~they finish off actuation with a heat gun if select hinges did not close, to (h)~form the target 3D geometry.}
  \Description{the workflow of OriStitch in visual form}
  \label{fig:teaser}
\end{teaserfigure}

%%\received{20 February 2007}
%%\received[revised]{12 March 2009}
%%\received[accepted]{5 June 2009}


\maketitle

\input{src/00_introduction.tex}
\input{src/01_contributions}
\input{src/02_relatedwork.tex}
\input{src/03_system.tex} 
\input{src/04_evaluation.tex}
\input{src/05_application.tex}
\input{src/06_discussion.tex}
\input{src/07_futurework.tex}

\input{src/08_conclusion.tex}
%\input{src/09_appendix.tex}

\begin{acks}
To Robert, for the bagels and explaining CMYK and color spaces.
\end{acks}

\bibliographystyle{ACM-Reference-Format}
\bibliography{references, thijs-papers, zekun}

%\input{src/09_appendix.tex}

\end{document}

%% file: src/00_introduction.tex
\section{Introduction}
Folding allows fabricating 3D structures from 2D sheets, which offers a range of advantages in fabrication workflows such as speeding up fabrication~\cite{an2018thermorph}, reducing material consumption~\cite{narumi2023inkjet}, flat packing objects for efficient transportation~\cite{tao2021morphing}, embedding sensors~\cite{wang2020morphingcircuit} and actuators~\cite{yao2015biologic} in otherwise passive structures. For the reasons above, as well as aesthetic effects, folding is a popular technique used when fabricating textiles.

A common form of industrial folding is machine pleating~\cite{zaki_innovation_2021}, while efficient and cheap, it only supports straight so-called "simple folds". More advanced folding enables rich textures such as the popular smocking pattern or textiles that maintain a 3D shape. These forms of folding require more elaborate fabrication, such as weaving~\cite{lind2019smocked} or embroidery~\cite{cane1985,gostelow2003,hall2008, luo_embroiderer_2023}. The resulting textures and shapes are interesting, but involve laborious manual processes. 

To achieve accessible workflows for folding of textiles, research focused on \textit{additive textile manufacturing}~\cite{keefe2022textile}. This includes techniques where folding patterns are added to fabrics either by printing~\cite{4d-textiles-additive-manufacturing-koch20214d}, or embroidery~\cite{4dorigami-embroidery-stoychev20174d}. The folds are actuated using shape memory alloys~\cite{seamless-seams-sara-nabil-DIS19}, actuated fibers~\cite{4DEmbroidery-Ayesha-Nabila-2023}, or even by controlling the fabrication of the fiber itself~\cite{forman_fiberobo_2023}. While these techniques are clever, they are hard to practically implement as they require manually laying out many individual hinges in machine instructions. \textit{Inkjet 4D Print}~\cite{narumi2023inkjet} is an automated, high-resolution folding process. The technique locally coats heat-shrinkable materials with ink to control folding angles. While the resulting geometry looks great, the resulting hinges unfold easily, and it only works on those shrinkable materials (and works best with plastic), limiting practical applicability in textile manufacturing.

We propose OriStitch, an end-to-end workflow to turn existing fabrics into self-folding structures through embroidery. Inherent to machine embroidery, this method is compatible with most existing off-the-shelf textiles, unlike weaving or knitting where a new textile is created from scratch, or relying on textiles that exhibit rare properties. Oristitch hinges tuck away material between faces when heat-actuated through the use of a shrinking thread, similar to \textit{4D Embroidery}~\cite{4DEmbroidery-Ayesha-Nabila-2023}. The specific design of OriStitch hinges allows them to fully close and lock in place. By laying out hinges on a 2D sheet and closing them, OriStitch allows converting flat-fabricated sheets into 3D structure. We develop an Oristitch software tool based on \textit{Origamizer}~\cite{Tachi2010:Origamizer} to automatically convert 3D meshes (OBJ) to stitch patterns, and compute an efficient division into hoops for fabrication. Exported files can directly be sent to laser cutters and embroidery machines. The resulting workflow allows creating 3D structures such as the cap shown in Figure 1 as well as other custom textiles like the vase cover and handbag shown in Figure~\ref{fig:applications}. OriStitch integrates well with other textile manipulation techniques, such as quilting, embroidery, or sewing, allowing it to fit in with existing production workflows without reducing the expressiveness of design.

Through technical validation, we find that OriStitch hinges fabricate and fold reliably using a range of materials (suede leather, cork, Neoprene, and felt). Our software tool successfully converted 26/28 models found in related papers~\cite{curveups-ruslan-guseinov-TOG17, narumi2023inkjet, inverse-design-inflatable-julian-panetta-TOG21,shrink-and-morph-David-Jourdan-TOG23} to machine instructions. With this contribution, we take a step to make the workflow of machine embroidery self-folding structures more accessible. The improved software support may allow fashion designers and discretionary users to design more elaborate patterns and advance the complexity of what they can make.

%% file: src/01_contributions.tex
\section{Contributions, Benefits, and Limitations}
This paper makes three contributions. (1)~We present a workflow, OriStitch, to create self-folding 3D geometry from flat fabrics. (2)~We present a design of stitch patterns to achieve fully closing hinges under heat treatment. And finally, (3)~we present a software tool to convert 3D meshes to fabrication instructions as well as an algorithm to efficiently compute hoop layouts. Combined, these contributions allow users to create a wide variety of shapes in a mostly automatic workflow. 

Inherent to the use of developable surfaces, this process is limited to 3D objects that are topologically disks. We demonstrate that with the cutting of the 3D shape and stitching together separate sheets, more complex topologies (such as topological spheres and tori) are attainable. Furthermore, to be able to place hinges, objects are subject to a resolution of edge lengths of a minimum of $8.4mm$. Our workflow uses high temperatures (>350$^{\circ}$F), as well as exposure to water; therefore, the used fabric has to be robust to these conditions.

%% file: src/02_relatedwork.tex
\section{Related Work}

This work builds on research in textile folding, software for textile fabrication, and techniques for self-folding.

\subsection{Textile Folding}
The fashion industry has a long history of creating folded textile structures, using a variety of techniques. 

\textit{Manual Processes.} Manual techniques, such as hand pleating, allow folding a range of materials. During hand pleating, fabric is placed between two halves of a cardboard mold, folded, and afterwards it is exposed to high temperatures. Another technique is hand-sewing continuous threads into textiles and pulling them together. Despite providing greater design freedom, these processes are labor-intensive and time-consuming, limiting their practicality.

\textit{Automated Processes.} A common automated folding process is machine pleating, where fabric is folded by two blunt knives, and then fixed at high temperatures. This technique is efficient for simple, repetitive patterns, but fails for complex folding. Knitted folding is typically achieved by varying the tension of knitted loops to create a structural disequilibrium of faces~\cite{pavko-cudenMultifunctionalFoldableKnitted2017}. \textit{Super Folds}~\cite{boon_samira_super_2015} are self folding fabrics through a combination of different yarns resulting in unbalanced tension. The same design studio (\textit{TextielLab}) has explored woven origami structures for deployment in space~\cite{sitnikova_self_2018}. An innovative approach, Steam Stretch~\cite{incISSEYMIYAKELdquo2014}, weaves combinations of cotton and polyester yarns. When steamed, the fabric morphs into 3D shapes and patterns. 

\textit{4D Embroidery Processes.} Embroidery is a technique to decorate materials with a needle and thread~\cite{higgin2022handbook}. This allows adjusting existing fabrics instead of weaving or knitting novel ones.~\citet{moore_precision_2018} demonstrate how to use embroidery to locally adjust the stiffness of materials. \textit{RandomPuff}~\cite{lin_randompuff_2021} creates tactile surfaces by applying heat shrinking patterns embroidered on existing fabrics. \citet{nabila4DEmbroideryImplementing2023a} propose an embroidery technique in which they embroider heat shrinkable patterns on lace-liked structures, resulting in self-folding 3D structures. \citet{4dorigami-embroidery-stoychev20174d} explore embroidery for 4D origami by simulating the behaviour of simple hinges. We build on this work by automatically converting 3D meshes to embroidery patterns, along with a special hinge design to enable shape-changing structures.

\subsection{Software Tools for Textile Fabrication}
In other domains of textile fabrication, several software tools have been built to support the design and manufacturing of 3D structures. Within computer knitting,
\textit{Knittable Stitch Meshes}~\cite{knittable-stitch-mesh-Kui-wu-TOG19} is a 3D modeling environment to design  3D knit instructions, \textit{Knitty}~\cite{igarashi_knitting_2008} allows knitting 3D animals, and \citet{Automatic-machine-knitting-Vidya-Narayanan-TOG18} propose a pipeline to convert 3D meshes to knitting instructions. \textit{KnitScript}~\cite{hofmann_knitscript_2023} introduces a scripting language, allowing others to build advanced interfaces and software to control the low-level functionality of knitting machines. Similarly, for automatic weaving, commercial software like \textit{ScotCAD}\footnote{\url{https://www.scotweave.com/products}} allows designing and editing complex patterns. \citet{harvey_weaving_2019} highlight the need for accessible workflows to allow for experimentation and democratization of digital weaving, they provide guidelines for the development of software.~\citet{wu_automatic_2020} a year later followed with an automatic workflow to synthesize 3D relief through weaving. \textit{WeaveCraft}~\cite{wuWeavecraftInteractiveDesign2020} is a full interactive editor for 3D weaving. And finally, \textit{EscapeLoom}~\cite{deshpande_escapeloom_2021} enhances digital weaving tools with new techniques to create advanced structures interwoven with other materials.

\citet{3D-smocking-Jing-Ren-TOG-2024} developed an algorithm to model and simulate 3D smocking patterns, allowing users to interactively explore results. \textit{Fabric Tessellation}~\cite{10.1145/3658151} proposed an inverse design tool to convert target 3D models to hand embroidered smocking patterns. OriStitch adds a new approach to convert 3D mesh models to machine embroidery patterns.

\subsection{Techniques for Self-Folding}
A wide variety of fabrication processes have recently been explored to achieve complex 3D shapes through 2D configurations. One of the most common self-folding techniques is to print thermoplastic composites using Fused Deposition Modeling (FDM) 3D Printers, also known as 4D printing~\cite{an2018thermorph}. Within textiles, similar methods create self-folding structures by printing on pre-stretched sheets in a process called \textit{textile additive manufacturing}~\cite{brenken2018fused}. \citet{keefe2022textile} presents a great overview of these techniques, highlighting embroidery as one such process. 

\citet{narumi2023inkjet} proposed a fabrication process using 4D inkjet printing to create self-folding 3D structures by locally coating special heat-shrinking materials with ink. This elegant and efficient process rapidly produces intricate structures, the fast folding under low forces and temperatures has a side-effect that unfolding is relatively easy too, making the technique excel at graphic rather than functional applications.

%% file: src/03_system.tex
\section{OriStitch}
\begin{figure}[h]
  \centering
  \includegraphics[width=0.62\linewidth]{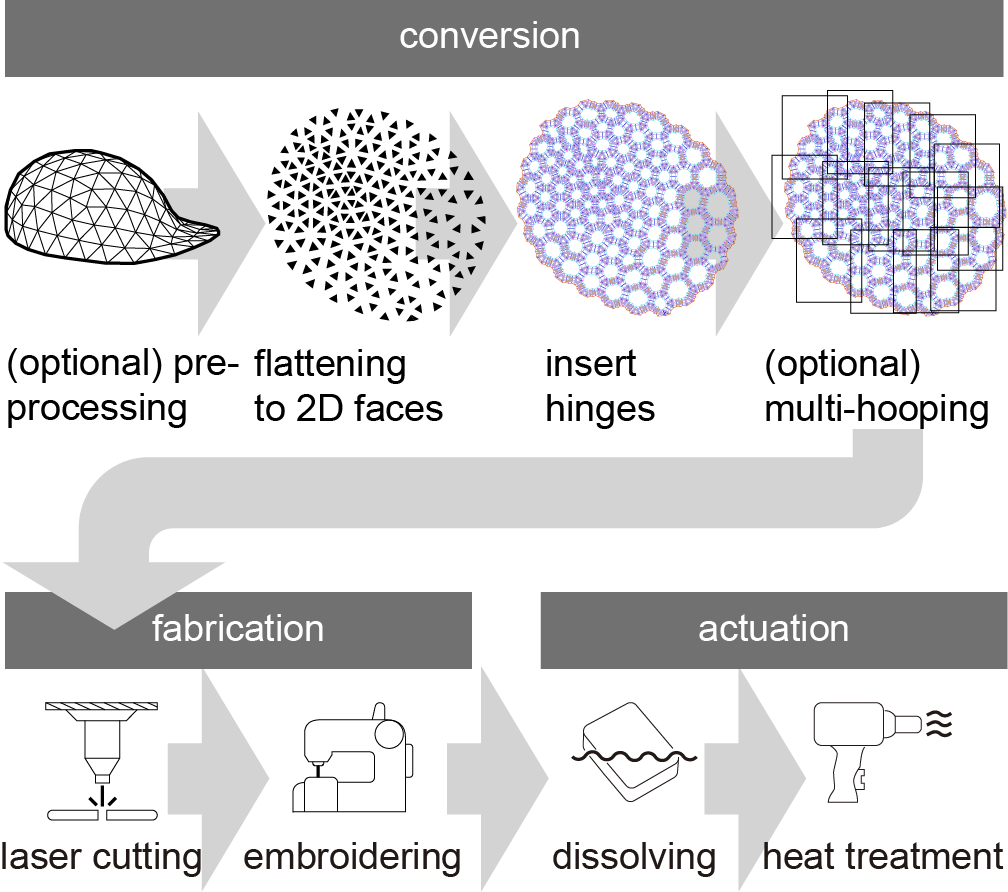}
  \caption{An overview of the OriStitch workflow.}
  \label{fig:system}
\end{figure}

The core of this work is the hinge design, OriStitch accompanies this with tools to place and distribute hinges based on an imported 3D mesh, while supporting multi-hooping. We first introduce the specific hinge and its mechanical design, and then introduce the computational tools to support the end-to-end fabrication workflow outlined in Figure~\ref{fig:system}.

\subsection{Stitch pattern design}
\begin{figure}[b]
  \centering
  \includegraphics[width=0.5\linewidth]{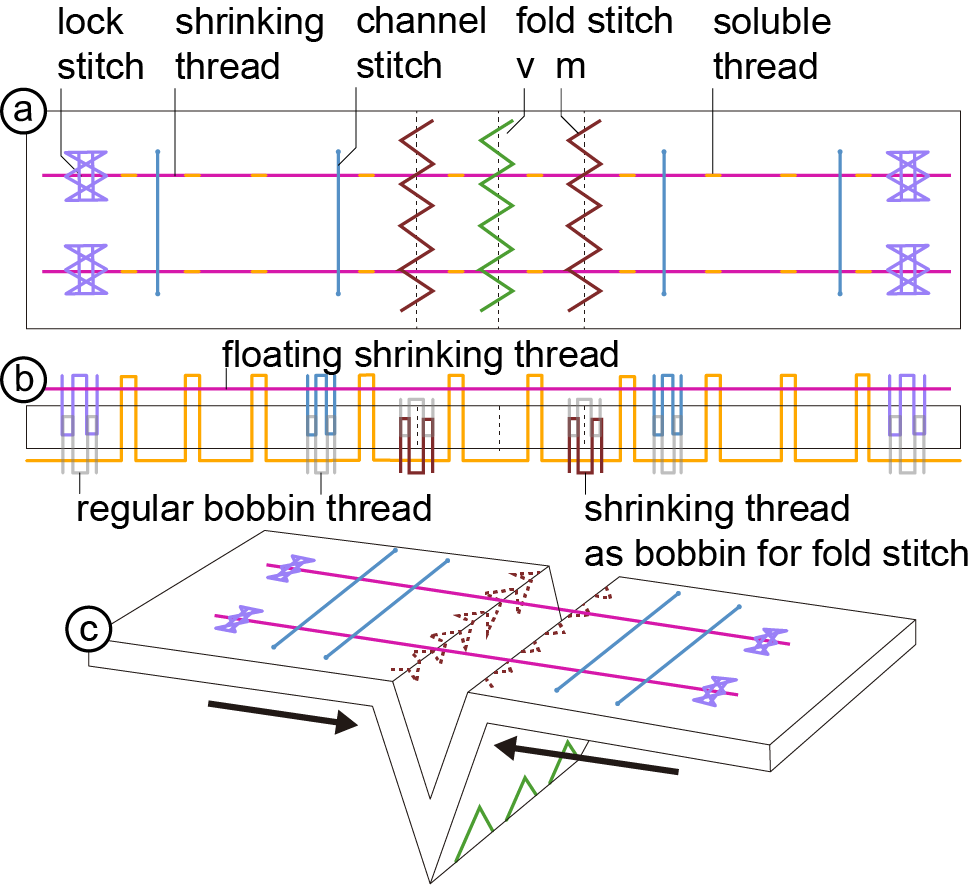}
  \caption{The basic hinge principle, (a)~from above, (b)~from the side, and (c)~when actuated.}
  \label{fig:single-hinge}
\end{figure}

Figure~\ref{fig:single-hinge} shows the elementary hinge and its actuation principle. The hinges fold along the valley fold on the reflecting axis, so that the space between faces is completely tucked behind the surface. As a result, the decomposed faces will be brought together to form the target 3D geometry. To accomplish this, the active thread is made of a special polyester (PE) thread (Chizimi)\footnote{https://thesewingplace.com/chizimi-shrinking-thread-300m} which shrinks by 30\% of its original length when exposed to heat (350$^{\circ}$F). The two ends of the heat-shrinking thread are fixed on a pair of adjacent faces by so-called \textit{lock stitches}. So that under heat treatment, as the active thread shrinks, the two faces will join. Each edge contains two parallel shrinking threads to uniformly close hinges. The so-called \textit{channel stitches} guide the shrinking thread on the faces. The \textit{fold stitches} are zigzag patterns with non-shrinking thread on top and shrinking threads on bottom (aka \textit{bobbin}) for mountain folds, and vice-versa for valley folds. As the thread shrinks, the edge pre-creases in the predefined orientation, making it easier for the shrinking thread to close the hinge. Finally, the orange, water-soluble thread holds the active threads in place during fabrication. 

\begin{figure}[h]
  \centering
  \includegraphics[width=0.5\linewidth]{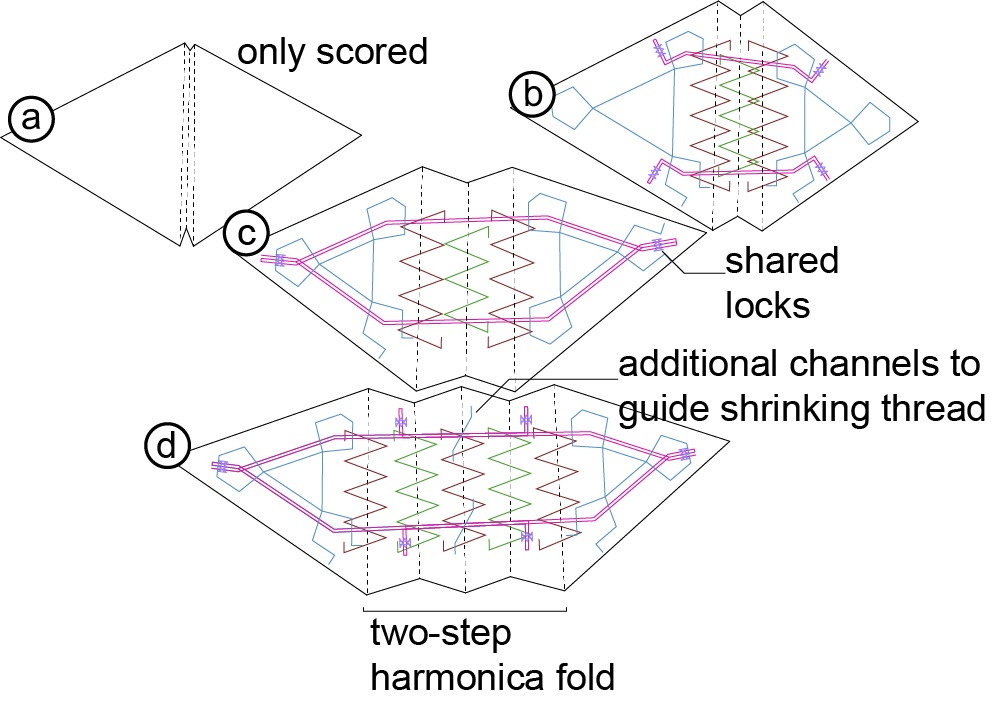}
  \caption{Hinge types of increasing width (a)~no hinge, just a fold between faces (b)~simple hinge (c)~extended by sharing lock stitches and wrapping around, (d)~or inserting harmonica-folds. Note also how all black lines are dashed to cut the pre-creased hinges.}
  \label{fig:hinge-types}
\end{figure}

\subsubsection{Variation of hinge types}
Between two faces, Oritstich inserts hinges, which tuck away excess material and ensure the faces meet up in the final 3D design. As a result the design of each hinge is dependent on the space between faces. The active thread shrinks by up to 30\%, so widening the hinge means more of the active thread has to be stitched to each face. To accommodate for this, OriStitch has four different hinge types displayed in Figure~\ref{fig:hinge-types}. (b) The basic hinge (type~1) places the threads as laid out in Figure~\ref{fig:single-hinge}. (c-d) to accommodate for wider hinges, Oristitch has hinges of (type~2) which wrap the shrinking thread around both faces to share lock stitches, and  (type~3)  wraps two paths of shrinking thread next to each other all the way around each of the faces while inserting an additional pair of hinges to produce a harmonica fold and a shrinking thread that spans all the way around the other hinge. (a) Given the optimized layout of faces, there are faces that are so close that they do not allow for hinges to be inserted; here, Oristitch simply creases the edge between the faces (type~0).

\begin{figure}[h]
  \centering
  \includegraphics[width=0.8\linewidth]{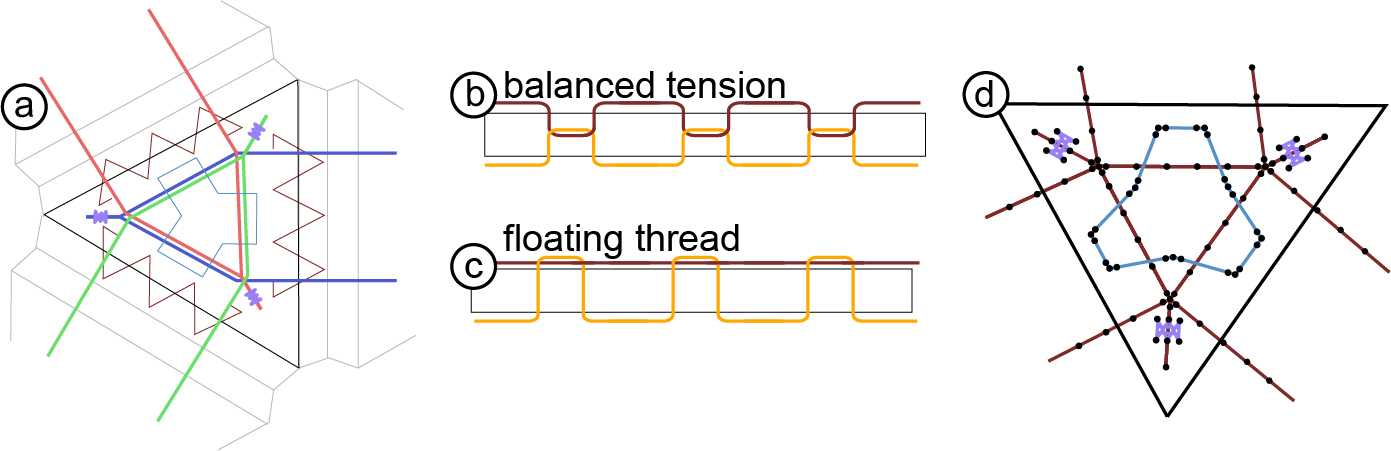}
  \caption{(a)The shrinking thread of neighboring faces (here red, green, and blue) overlaps on each face to share channel and lock stitches. (b)~Typically, the bobbin thread pulls the top thread through the fabric for a secure fit at each needle point. (c)~To keep the shrinking thread flexible, OriStitch does not pull it into the fabric (d)~OriStitch adds needle points to distribute tension on shrinking threads and reinforce the corners.}
  \label{fig:techfigs}
\end{figure}

As illustrated in Figure~\ref{fig:techfigs}a, each face has up to three neighboring faces, so they contain the shrinking patterns for up to three hinges. The fold stitches bend along the edges. This avoids intersections between the active threads, allowing them to share channel stitches and lock stitches. Besides simplifying the layout, sharing channel, and lock stitches speeds up fabrication.

\subsubsection{Hinge type placement algorithm}

The placement of each hinge is determined by a two-stage classification algorithm. This method evaluates the absolute width of a hinge and its width relative to the available folding space, which is a robust indicator of whether the material can physically accommodate the fold. 

\begin{figure}[h]
  \centering
  \includegraphics[width=0.3\linewidth]{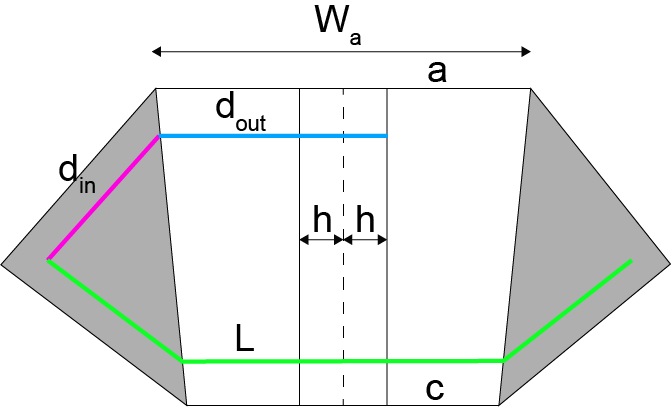}
  \caption{A simplified hinge pair with all technical dimensions labeled. Variables \( d_{\text{in}} \) and \( d_{\text{out}} \) are only relevant in type 3 hinges.} 
  \label{fig:specs}
\end{figure}

First, each side of a hinge (a and c) is independently assigned a Gap Type, $T(w, L)$, an integer score from 0 (No Gap) to 3 (wide). This score is a function of the hinge width $w$ and the available "shrinking path" length $L$. Figure~\ref{fig:specs} visually depicts these parameters in a prototypical hinge pair. The function prioritizes hinge sides that are already very narrow ($w < 2.0$) and classifies no-gap type. For wider hinges, it calculates the critical ratio of width to available space, $\frac{w}{L}$, and assigns a score based on predefined design thresholds ($R_{narrow}=0.25$, $R_{medium}=0.3$, $G_{close} = 2.0$):

\[
T(w, L) = 
\begin{cases} 
0 & \text{if } w < G_{close} \text{ (No Gap)} \\
1 & \text{if } w \ge G_{close} \text{ and } w/L \le R_{narrow} \text{ (Narrow)} \\
2 & \text{if } w \ge G_{close} \text{ and } R_{narrow} < w/L \le R_{medium} \text{ (Medium)} \\
3 & \text{if } w \ge G_{close} \text{ and } w/L > R_{medium} \text{ (Wide)}
\end{cases}
\]

Second, the hinge type is determined from the combination of the two sides' gap types. By analyzing the sorted pair of scores, $P = (\min(T_a, T_c), \max(T_a, T_c))$, the algorithm applies a set of rules that reflect practical manufacturing constraints. For example, a hinge is only classified as the type 0 if both sides are have no gap ($P = (0, 0)$), while a combination like $(1, 2)$ is still acceptable and classified as type 1. This two-step process provides a nuanced and reliable assessment of each hinge's feasibility.

\begin{itemize}
    \item \textbf{Type 0 (No hinge)}: If $P = (0, 0)$
    \item \textbf{Type 1 (Narrow)}: If $P \in \{(0, 1), (1, 1), (1, 2)\}$
    \item \textbf{Type 2 (Medium)}: If $P = (2, 2)$
    \item \textbf{Type 3 (Wide)}: If $P \in \{(2, 3), (3, 3)\}$
    \item \textbf{Error}: For all other combinations.
\end{itemize}
 
The way the Origamizer algorithm spaces out the faces from one another makes error cases of (0,2),(0,3), and (1,3) cases extremely rare. In our technical evaluation, we have encountered none of them. These would result in folding angles that our hinge design does not support. Users receive a warning when this happens. 

\subsubsection{constraints of the type 3 design}

The active thread used in this paper shrinks by 30\%, this constraints the design of hinges. For the long hinge design (type 3), the amount of material to fold into the harmonica fold depends on the total shrinkage. Here we lay out the way these constraints impact the design to get a sense for how to adjust to potential other threads. The basic premise is simple: we have an active thread, of length $T$, and a hinge that tucks away material of height $h$. 

Let:
\begin{itemize}
    \item \( W_a \): the distance between anchor points on the two triangles
    \item \( d_{\text{in}} \): the portion of the thread embedded inside the triangle
    \item \( d_{\text{out}} \): the portion of the thread extending outside the triangle
    \item \( T = d_{\text{in}} + d_{\text{out}} \): the total length of the thread used for this pair
\end{itemize}

To achieve the desired folding, the total shrinkage of the thread must equal \( 2h \). The thread we use has a shrinkage factor \( S_f = 30\% \). From this geometric constraint: 

\begin{align}
0.3 \, T &= 2h \nonumber \\
T &= \frac{2h}{0.3} = \frac{20}{3} h
\label{eq:ht}
\end{align}

Therefore:

\begin{equation}
d_{\text{out}} = \frac{20}{3} h - d_{\text{in}}
\label{eq:dout_h}
\end{equation}

We enforce two fabrication constraints:

1. To prevent the thread from extending too far and intruding into the opposite triangle, we ensure:

\begin{equation}
d_{\text{out}} < \frac{3}{4} W_a
\label{eq:dout_ub}
\end{equation}

2. To ensure the thread sufficiently spans the hinge region and supports folding behavior, we require:

\begin{equation}
d_{\text{out}} > \frac{1}{2} W_a + h
\label{eq:dout_lb}
\end{equation}

Substituting \ref{eq:ht} into \ref{eq:dout_ub} \ref{eq:dout_lb}  gives the constraint bounds on \( h \):

\begin{equation}
\frac{1}{2} W_a + h < \frac{20}{3} h - d_{\text{in}} < \frac{3}{4} W_a
\label{eq:h_bound}
\end{equation}

Solving for \( h \) from both sides, we get:

\begin{align}
\text{From the lower bound:} &\quad \frac{20}{3} h - d_{\text{in}} > \frac{1}{2} W_a + h 
\Rightarrow h > \frac{3}{17} \left( \frac{1}{2} W_a + d_{\text{in}} \right) 
\label{eq:h_lower_bound} \\
\text{From the upper bound:} &\quad \frac{20}{3} h - d_{\text{in}} < \frac{3}{4} W_a 
\Rightarrow h < \frac{3}{20} \left( \frac{3}{4} W_a + d_{\text{in}} \right) 
\label{eq:h_upper_bound}
\end{align}

Therefore, the hinge height \( h \) must satisfy:

\begin{equation}
\frac{3}{17} \left( \frac{1}{2} W_a + d_{\text{in}} \right) < h < \frac{3}{20} \left( \frac{3}{4} W_a + d_{\text{in}} \right)
\label{eq:final_bound}
\end{equation}

This expression provides a valid design window for choosing \( h \), ensuring both geometric feasibility and fabrication compliance.
In our implementation, we select the hinge height \( h \) as the midpoint of the feasible interval to balance thread coverage and prevent interference with adjacent triangles. Specifically, we choose:

\begin{align}
h &= \frac{1}{2} \left[
\frac{3}{17} \left( \frac{1}{2} W_a + d_{\text{in}} \right) +
\frac{3}{20} \left( \frac{3}{4} W_a + d_{\text{in}} \right)
\right] \notag \\
&= \frac{273}{2720} W_a + \frac{111}{680} d_{\text{in}}  \label{eq:h_midpoint} \\
&\approx 0.1004 \, W_a + 0.1632 \, d_{\text{in}} \label{eq:h_midpointap}
\end{align}

By grounding the hinge parameters in both geometric constraints and thread material behavior, our method ensures that each hinge reliably contracts to the target folding height after shrinking, preserving the intended mechanical behavior and foldability.

Our next question is, by using this method of selection, will this satisfy all cases? Or is there any restriction we have here? The restriction is that the folding hinge h can not be too small $<2mm$, otherwise folding performance will not be guaranteed, though a smaller $h$ will support a longer hinge. Given our hinge design in \ref{eq:h_midpoint}, we impose the constraint:

\begin{equation}
h = \frac{273}{2720} W_a + \frac{111}{680} d_{\text{in}} \geq 2 mm
\label{eq:h_constrin}
\end{equation}

Multiplying both sides by 2720 yields:

\begin{equation}
273 \, W_a + 444 \, d_{\text{in}} \geq 5440mm
\label{eq:h_constrin_result}
\end{equation}

This inequality defines the minimal combination of anchor width \( W_a \) and inner thread length \( d_{\text{in}} \) required to maintain a hinge height of at least 2 mm. For usage with other threads we recommend characterizing the precise shrinkage and adjusting the parameters of these constraints accordingly.

\subsubsection{Needle points}

OriStitch' hinge design requires careful placement of needle points, the stitches where bobbin threads (from below) connect with the threads on top. As illustrated in Figure~\ref{fig:techfigs}b, in conventional embroidery, designers prioritize the top side of the fabric for aesthetic reasons, they provide extra tension on the bobbin thread, pulling it through the sheet. OriStitch cannot afford that because the active threads need to smoothly adjust their length. OriStitch adds additional needle points to distribute the tension across the thread, as shown in Figure~\ref{fig:techfigs}c. This is specifically important at corners, where the thread on top makes a turn and would tear out the bobbin threads. OriStitch adds additional needle points at all corners and has less on the straight paths, ensuring both reliable fabrication and actuation.

\subsubsection{Resolution}
\begin{figure}[h]
  \centering
  \includegraphics[width=0.7\linewidth]{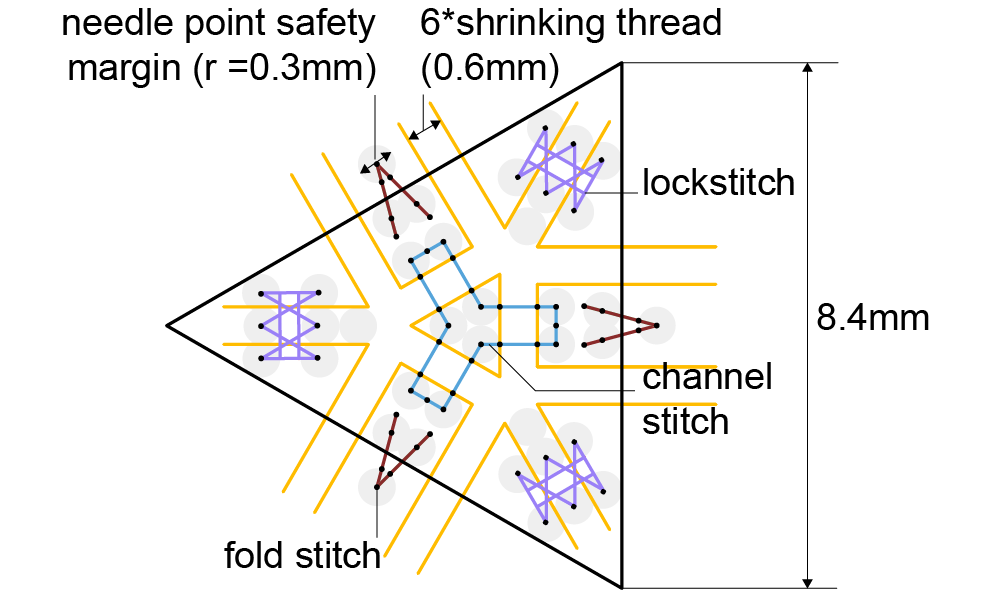}
  \caption{The minimum size of a face to convert to OriStitch hinges is 8.4mm, constrained by the (machine-specific) size of the needle point safety margin.}
  \label{fig:resolution}
\end{figure}

OriStitch is subject to a minimum size for each face to allow sufficient space to place all stitches. The minimal size of faces to be converted to OriStitch hinges has an edge length of 8.4mm and is predominantly constrained by embroidery machine limits. Typical machines have a needle point safety margin (distance between points) of 0.3mm, as shown in Figure~\ref{fig:resolution}, this is the main constraint for the size of faces in OriStitch. The channel stitch in the center is reduced to the minimal size, each corner requires three needle points, so on both sides of the active thread path there are exactly three points. The active thread path contains up to 6 parallel threads, which, in order for the threads to stay flexible, requires 0.6mm of width. The final constraining factor is the locks stitch design, which again for safety margin reasons cannot be further reduced. The main conclusion is that the resolution really is machine-specific when it comes to the safety margins. That said, this assumes a dense fabric (leather, felt, or materials with tightly woven structures), if the fabric has a structure with wider gaps (typical for loosely woven or knitted structures), we recommend increasing the minimum edge length in the model past this resolution before converting to stitch patterns. The Oristitch editor warns users when their OBJ file contains triangles that have edges below this minimum resolution, in this case users will have to re-triangulate their file.

\subsection{Fabrication process} 
\begin{figure}[b]
  \centering
  \includegraphics[width=0.7\linewidth]{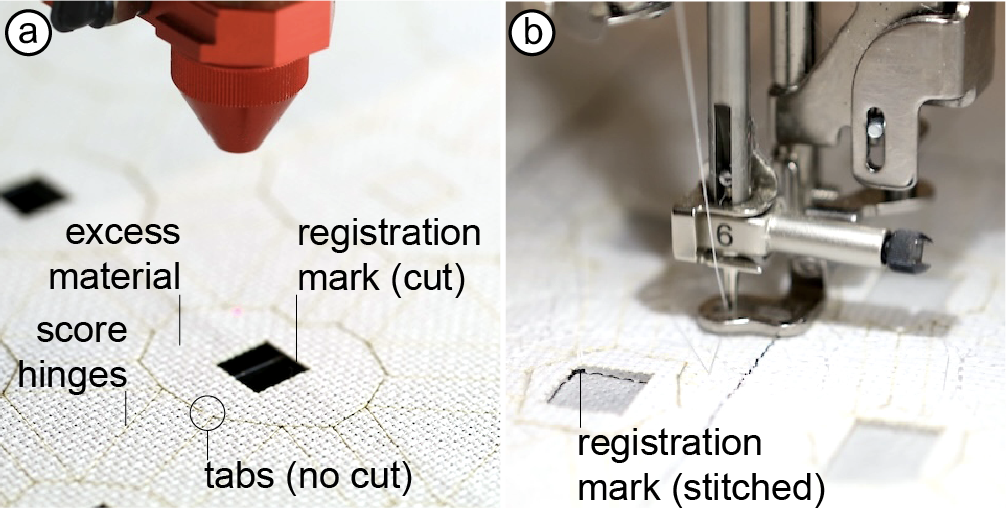}
  \caption{The fabrication workflow involves (a)~laser cutting to score hinges and cut away excess material (b)~and then embroidery. The laser cutting file contains registration marks to align the embroidery process afterwards.}
  \label{fig:fabrication}
\end{figure}

The fabrication process of the 2D sheets consists of two phases, first laser cutting and then followed by embroidery, as shown in Figure~\ref{fig:fabrication}.

\subsubsection{Materials}
The core material we used in this paper is a special polyester (PE) thread (Chizimi) that shrinks around 30\% under heat treatment\footnote{https://thesewingplace.com/chizimi-shrinking-thread-300m/}. We use water-soluble material\footnote{\label{note-soluble}https://www.fjx.co.jp/english/catalog/household/water.html} to temporarily hold other threads in place before actuation. Most of our testing has been done with "Aida cloth, 18 count" fabric\footnote{https://www.michaels.com/product/10357725}, but we find in the technical evaluation that the hinges continue to work with a range of fabrics. That said, caution should be placed on the selection of fabrics. During the actuation process both exposure to wetness and high temperatures limit the range of effective materials. We also found in our evaluation that the materials need to have some stiffness to work reliably, otherwise the hinges fail to fold as expected. OriStitch comes with a calibration patch shown in Figure~\ref{fig:calibration}, which we recommend to use for calibrating to material properties as well as actuation temperature.

\subsubsection{laser cutting}

Next to machine-embroidery, OriStitch's workflow makes use of laser cutting. As most fabrics are relatively flexible, they require scoring or pre-creasing of the edges to create well-defined folds with OriStitch. To reduce the force required to close the hinges and to create well-defined edges, mountain and valley folds cut along most of the edges, with only the top, middle, and bottom sections connected as demonstrated in Figure~\ref{fig:hinge-types},
Oristitch cuts away excess material between vertices to make it easier to fold. While the cutting step is not strictly required for the pattern to fold (as the original layout as generated by \textit{Origamizer}~\cite{Tachi2010:Origamizer} folds paper without cutting), it simplifies the 2D layout as there is no need for additional patterns to fold the material away (aka "vertex tucking molecules"), which in turn reduces the forces required to close each hinge. 

As shown in Figure~\ref{fig:fabrication}, OriStitch adds little tabs in the cutout areas to keep the sheet together; this minimizes the risk of the fabric folding on top of itself during embroidery and it keeps the tension in the sheet more even. Within the holes, OriStitch cuts small squares that are used for registration in the multi-hooping process. (b)~OriStitch embroiders the same squares on the stabilizer sheet before attaching the fabric to the hoop. When attaching the fabric, these rectangles help align both sheets. Alternatively to laser cutting, some embroidery machines use cutting needles which can be used to remove the excess material. OriStitch exports both the laser cutting plan and stitch patterns as vector graphics (SVG). 

\subsubsection{Dissolving support threads}
As illustrated in Figure~\ref{fig:single-hinge}b, the heat-actuated thread is held in place with a water-soluble bobbin thread\textsuperscript{\ref{note-soluble}}. During embroidery, something needs to hold the actuated thread in place, which is where this specific bobbin thread comes in. As the figure shows, the bobbin thread is very low in tension so as not to pull the shrinking thread through the fabric (as opposed to the other grey bobbin threads for lock and channel stitches, which are pulled in place). When exposed to water, these bobbin threads dissolve, allowing the heat-actuated thread to shrink unconstrained. 

\begin{figure}[h]
  \centering
  \includegraphics[width=0.7\linewidth]{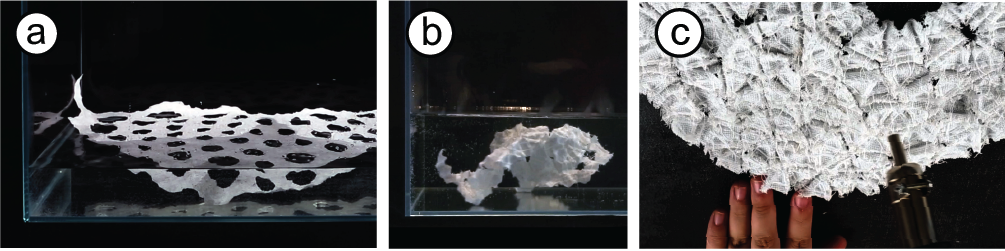}
  \caption{update with new picture: The last steps of the workflow, (a)~dissolving the soluble thread, (b)~pre-creasing the hinges in boiled water, and (c)~heat-actuating the remaining hinges.}
  \label{fig:actuation}
\end{figure}

\subsection{Heat-actuating the hinges into their final shape}

To actuate the pattern, we propose dipping the sheet in boiling water, if needed, followed by post-processing with a heat gun as shown in Figure~\ref{fig:actuation}. The combination of hot water and heat gut treatment enables a relatively fast and reliable actuation with minimal manual labor. The hot water pre-creases all hinges to ±70\% of their folded state in a matter of seconds. Models consisting of a small number of triangles completely fold in boiling water because the lever each fold pulls is shorter. The focused distribution of heat with the heat gun ensures that each hinge closes one at a time instead of all at once (in water).

\begin{figure}[h]
  \centering
  \includegraphics[width=0.7\linewidth]{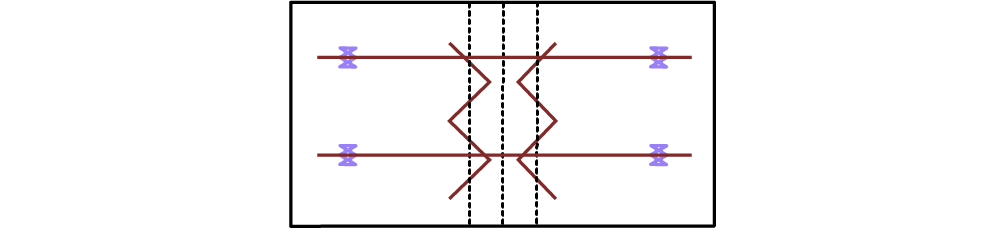}
  \caption{The calibration gauge is a minimal hinge; we recommend increasing the actuation temperature from 300$^{\circ}$F until closed.}
  \label{fig:calibration}
\end{figure}

We attempted to use a convection oven as well, which did actuate all hinges in parallel, however simultaneous closing caused hinges to pull on one another at the same time, and the high temperature made the actuated threads fragile, by actuating sequentially, we circumvent that problem albeit a bigger manual effort. Some hinges require more tension than others because of their width and neighboring faces, therefore the manual process allows actuating each hinge until it fully closes. The shrinking conditions depend on various factors, including the temperature, the amount of air flow, the distance between the hot air and the threads, and the duration of treatment. Therefore, instead of proving a specific temperature, we recommend using a simple gauge shown in Figure~\ref{fig:calibration} to calibrate the heating conditions for the used heating mechanism and materials. 

\subsection{Computational tools to support the OriStitch workflow} 
Finally, to give users a chance to benefit from the Oristitch hinge design to produce interesting 3D structures, we provide computational tools to (1)~convert a target 3D mesh into machine instructions and (2)~support multi-hooping, a process that allows fabricating larger sheets on small embroidery machines. 

\subsubsection{Converting 3D meshes to machine instructions}
\begin{figure}[h]
    \centering
    \includegraphics[width=1\linewidth]{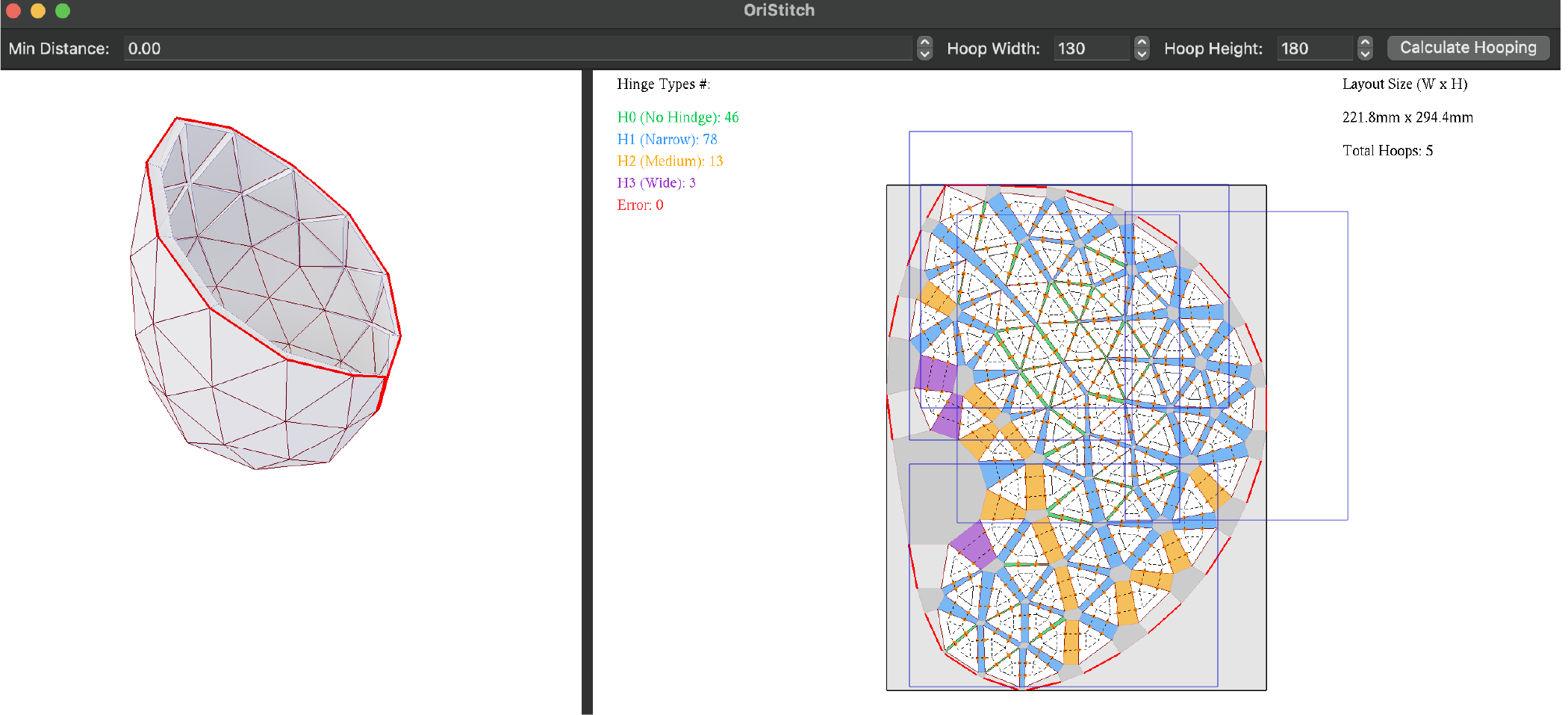}
    \caption{Users load the OBJ model (left), which produces the stitch pattern on the right. Users see these at the same time, and can move the triangles in the 2D view (which highlights them in the 3D model) to achieve their desired layout. The interface feeds back hinge types and work size, and when hitting the "calculate hoping" button, it computes the number of hoops for multi-hooping and overlays them in the visualization.}
    \label{fig:ui}
\end{figure}

Oristitch's interface is a simple GUI based on Tachi's \textit{Origamizer}~\cite{Tachi2010:Origamizer}. This tool converts a given mesh into a 2D configuration of faces, built upon the \emph{Origamizer}~\cite{Tachi2010:Origamizer} algorithm, which achieves a similar objective in the context of paper folding. The algorithm transforms a disk-equivalent mesh into 2D by calculating each face's rigid transformation and introducing connecting creases. These creases tuck away excess material either at the edges or vertices. The Oristitch interface (shown in Figure~\ref{fig:ui}) allows users to tune the resulting crease patterns interactively, includes machine-specific constraints, subdivides the pattern for multi-hooping, and exports fabrication-ready machine instructions for laser cutters and embroidery machines.

Oristitch is an extension of the two-phase mapping algorithm for origamizing polyhedral surfaces, as introduced in \textit{Origamizer}~\cite{Tachi2010:Origamizer}. We preserve the original framework's core strength in ensuring geometric foldability and focus our contribution on the \textit{Translational Mapping} phase, where the final positions of the unfolded polygons are determined.

The original algorithm in this phase uses an iterative optimization process to find a valid, non-overlapping layout. Its primary objective is to minimize a penalty energy, $E_w$, derived from a set of geometric and topological constraints. We augment this process by introducing a new, tunable penalty term that allows a designer to enforce a user-defined minimum separation distance, a critical constraint for physical fabrication as machine and material properties enforce their own minimum separation for successful fabrication, and users may have aesthetic preference to further increase the distance at the trade-off of a larger sheet to fabricate. Oristitch' user interface shows users the size, and number of each hinge type, which updates interactively as the user manipulates the stitch pattern, as shown in Figure~\ref{fig:ui}.

Oristitch accomplishes this by modifying the objective function that the optimizer minimizes. The total penalty energy, $E_{total}$, is now a weighted sum of the original geometric penalty and our new distance penalty:
\begin{equation}
    E_{total} = E_{geom} + \lambda \cdot E_{dist}
\end{equation}
Here, $E_{geom}$ represents the original penalties from Tachi's method. The new term, $E_{dist}$, enforces our minimum distance rule, $d_{min}$, and $\lambda$ is a user-defined weight to control its influence. The penalty $E_{dist}$ is formulated as a one-sided quadratic function; it is zero if the separation between two parts is greater than or equal to $d_{min}$ and grows quadratically as the separation distance drops below this threshold. This creates a strong, stable repulsive force that enforces the spacing constraint.

Crucially, this penalty is applied to the algorithm's internal proxy distance metric. Oristitch leverages this existing proxy because, while not strictly identical to the Euclidean distance, it is a smooth and computationally stable function of the layout parameters. The gradient of this proxy metric is continuous, which is essential for the stability of the numerical solver. Using this well-behaved metric ensures that the new penalty integrates seamlessly into the original gradient-based optimization framework, preserving the robustness and convergence properties of the foundational algorithm, while respecting fabrication constraints and user preferences.

Given the 2D planar layout of faces, OriStitch places the cutting paths and stitch patterns based on the four different types of hinges outlined in the previous section.

\subsubsection{Multi-Hooping} 

A single embroidery hoop may not suffice to contain the entire unfolded layout. While large industrial machines should be able to cover most basic stitch pattens, there are limitations for smaller (household) machines, or for very large patterns. A common strategy in embroidery to address this is called "multi-hooping". Operators effectively hoop a large sheet of fabric into a smaller hoop, they stitch the first pattern, then "rehoop" by placing the hoop on a new part of the sheet, typically slightly overlapping with the previously embroidered pattern. The graph nature of Oristitch folding patterns (each face connects to the others by the active thread) makes it non-trivial to naively divide the total design into individual hoops.

\begin{figure}[h]
  \centering
  \includegraphics[width=0.7\linewidth]{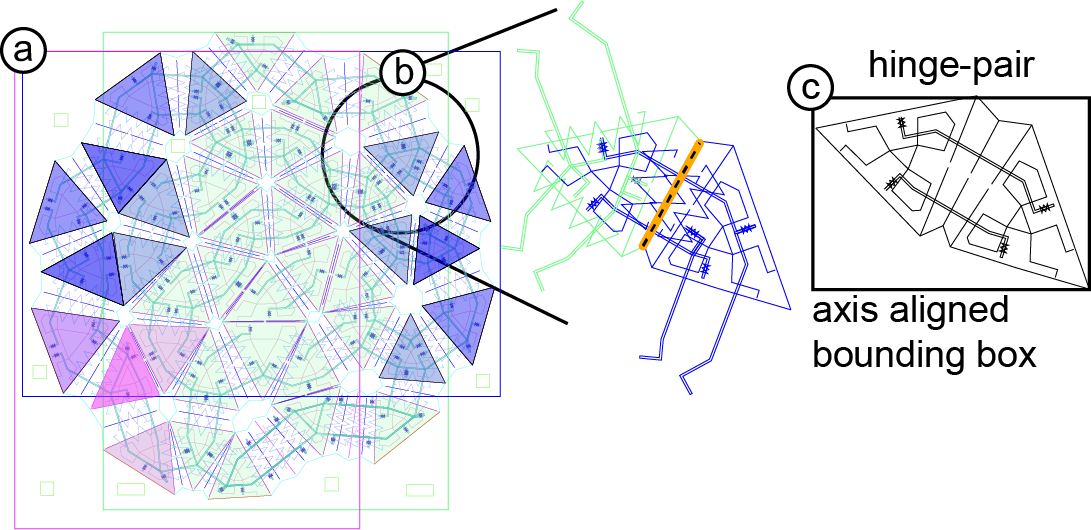}
  \caption{OriStitch supports multi-hooping to make the design work reliably on user's specific machines. (b)~It takes care of the fabrication sequence of the hoops to ensure hinge functionality. (c)~to make computation tractable, it represents hinge-pairs by their axis aligned bounding box (AABB).}
  \label{fig:multi-hooping}
\end{figure}

Oristitch therefore strategically subdivides the stitch pattern based on the machine-specific hoop size as shown in Figure~\ref{fig:multi-hooping}. Unlike typical multi-hooping techniques, our technique takes into account the specific sequence of fabrication steps for our hinges to work, and the division within hinges, Figure~\ref{fig:multi-hooping}b shows how the pink (hoop 2) face is laid out first, and then in green (hoop 3) the active thread is looped in and the lock and channel stitches are added. When users export their fabrication files, the resulting stitch pattern is subdivided into separate hoops to fabricate. 

To subdivide the stitch pattern, Oristitch computes the minimal number of vertical or horizontal hoop placements required to fully cover all hinges and their corresponding faces (we call "hinge-pairs").

\paragraph{Boundary Extraction.} 
Each valid gluing edge in the polyhedral model defines a hinge-pair. Instead of computing the precise shape boundary, we use the axis-aligned bounding box (AABB) of these vertices to represent its extent. That is, for each unit we store its $x_{\min}$, $y_{\min}$, $x_{\max}$, $y_{\max}$. This representation will not lose precision because the we assume axis-aligned hoops (slightly more compact hooping may be achieved by allowing for rotation, but as we will see later, comes at a significant computational cost).

\paragraph{Problem Formulation.}
Each hinge-pair is bounded by an axis-aligned rectangular bounding box $\mathsf{AABB}_i$, and the set of all such boxes is denoted by $\mathcal{H} = \{\mathsf{AABB}_1, \mathsf{AABB}_2, \dots\}$. The objective is to cover all rectangles in $\mathcal{H}$ using the minimum number of embroidery hoops, each modeled as a rectangle of fixed width $w$ and height $h$ (defined by users depending on their available hoop size).

This is a \emph{set covering problem}, known to be NP-hard. Therefore, we propose a practical greedy heuristic that efficiently places hoops to maximize coverage. Because of the fabrication sequence constraints, each face has to be fully contained in a hoop; taking the AABB of hinge-pairs will safeguard that property. The cyclical nature of the stitch pattern (all face edges except for the boundary are connected to other edges) ensures that there is always required overlap between neighboring hoops to fabricate the stitches shared among hinges.

\paragraph{Greedy Multi-hooping algorithm.}
The algorithm proceeds iteratively as follows:

\begin{enumerate}
    \item \textbf{Anchor Candidate Extraction.}
    Compute the union of all bounding boxes in the hinge-pair set $\mathcal{H}$, where each unit is represented by an axis-aligned rectangle $\mathsf{AABB}_i = [x_{\min}^i, x_{\max}^i] \times [y_{\min}^i, y_{\max}^i]$. 
    To identify a discrete set of candidate anchor positions for hoop placement, we collect all distinct $x$- and $y$-coordinates from the bounding box edges:
    
    \begin{equation}
    \mathcal{X} = \bigcup_{\mathsf{AABB}_i \in \mathcal{H}} \{x_{\min}^i\}, \quad
    \mathcal{Y} = \bigcup_{\mathsf{AABB}_i \in \mathcal{H}} \{y_{\min}^i\}.
    \label{eq:hoopxy}
    \end{equation}
    
    The Cartesian product $\mathcal{X} \times \mathcal{Y}$ defines a discrete set of anchor candidates, where each $(x, y)$ corresponds to a potential lower-left corner for placing a hoop.
    
    \item \textbf{Hoop Simulation.} For each $(x, y) \in \mathcal{X} \times \mathcal{Y}$, simulate placing a hoop at that anchor with two orientations:
    \begin{itemize}
        \item \textbf{Horizontal:} hoop size $w \times h$, bottom-left corner at $(x, y)$;
        \item \textbf{Vertical:} hoop size $h \times w$, bottom-left corner at $(x, y)$.
    \end{itemize}
    Count how many remaining boxes in $\mathcal{H}$ are fully contained within each simulated hoop.

    \item \textbf{Greedy Selection.} Select the hoop placement (position and orientation) that covers the largest number of bounding boxes. If no placement covers any remaining box, the algorithm terminates with failure.

    \item \textbf{Update.} Remove all covered boxes from $\mathcal{H}$ and repeat from step 1 until $\mathcal{H}$ is empty.
\end{enumerate}

\paragraph{Runtime Analysis.}
Let $N$ be the number of Hinge-pairs. Each is represented by an axis-aligned bounding box.

In each iteration, we collect at most $N$ distinct $x$ and $y$ values from all bounding boxes, yielding $\mathcal{O}(N^2)$ candidate anchor positions. For each position, we evaluate two hoop orientations and check containment against $\mathcal{O}(N)$ boxes, resulting in $\mathcal{O}(N^3)$ work per iteration.

As each iteration removes at least one unit, there are at most $N$ iterations. The total worst-case runtime is:
$$
\mathcal{O}(N) \times \mathcal{O}(N^3) = \mathcal{O}(N^4).
$$
Thus, the algorithm runs in polynomial time.

% \begin{figure}[h]
%     \centering
%     \includegraphics[width=0.9\linewidth]{figures/load-develop-model.png}
%     % 
%     \fbox{\parbox[c][4cm][c]{0.9\linewidth}{\centering Placeholder for UI showing real-time updates of size and hinge type count.}}
%     \caption{User interface for model loading and development. Users can apply distance constraints and manually adjust the layout. Real-time statistics are shown for hinge types and unfolded dimensions.}
%     \label{fig:load-develop-model}
% \end{figure}

\subsubsection{Export Stitch Pattern}
Once the unfolded model is developed and the hoop placement is completed, users export a machine-ready SVG file for fabrication. The generated SVG is structured in a hierarchical grouping format to reflect the embroidery and laser cutting processes.

\begin{figure}[h]
    \centering
    \includegraphics[width=1\linewidth]{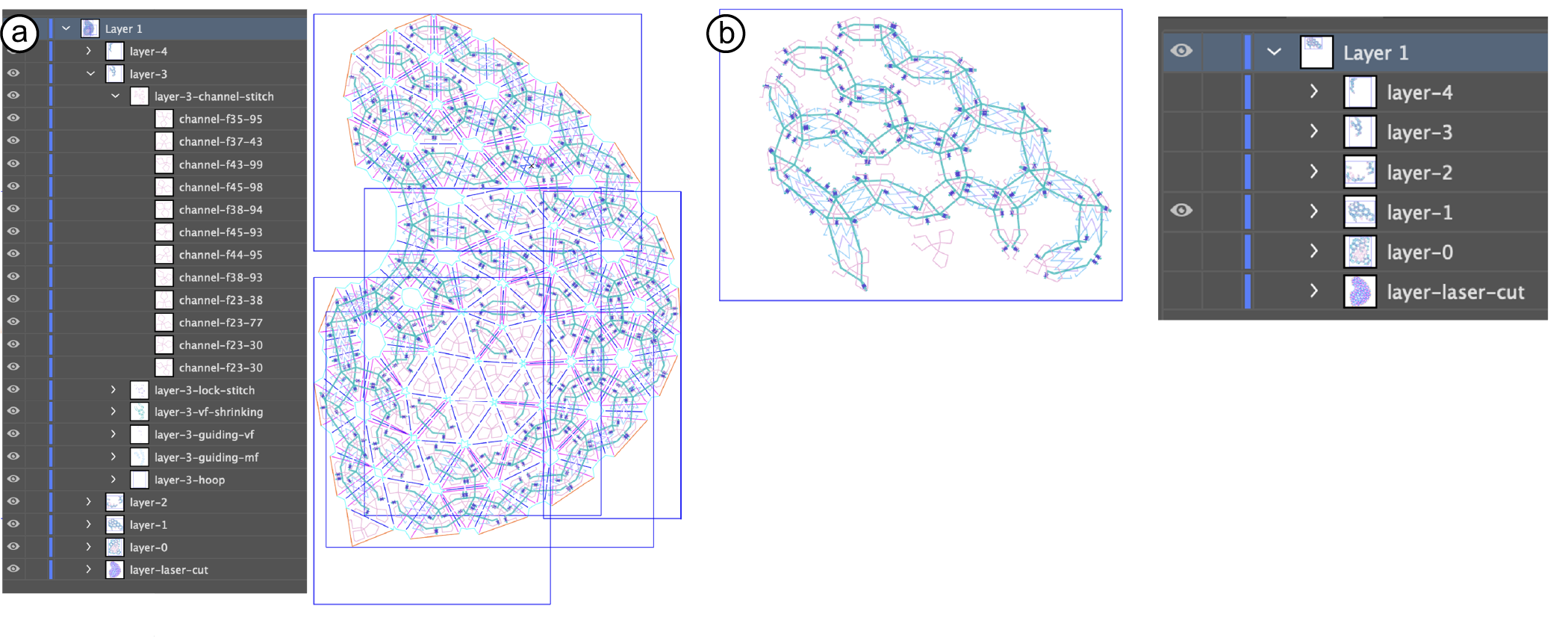}
    \caption{(a) Hierarchical structure of the exported SVG file: top-level hoop layers, stitch-type subgroups, and path ordering.(b) zoomed in to an individual hoop.}
    \label{fig:svg-export}
\end{figure}

As shown in Figure ~\ref{fig:svg-export}, at the top level, each group corresponds to either a specific hoop placement or the laser-cut layer. Within each hoop group, subgroups are used to differentiate between various types of stitching patterns, such as mountain folds, valley folds, locking stitches, and others. Furthermore, within each stitching pattern subgroup, the individual paths are sorted using a greedy algorithm. We use a greedy distance-based algorithm that iteratively selects the next nearest stitching path to minimize travel distance and reduce non-stitching jumps during embroidery. This sorting approach helps reduce excessive jumping between stitches and facilitates a more efficient machine operation.

This structured output allows direct integration with embroidery machines and supports both laser cutting and stitching operations. The layering ensures that the fabrication order is preserved and interpretable.

%% file: src/04_evaluation.tex
\section{Evaluation}

To evaluate our contributions, we conducted two technical evaluations. First, we evaluated the success rate of our shape conversion tool, and secondly, we evaluated our fabrication workflow and hinge behavior on a range of different materials. 

\subsection{OriStitch automatically converted 26/28 mesh models to stitch patterns}
To evaluate the mesh conversion, we used OriStitch to replicate mesh models found in related research papers. We measured the success rate, the types of hinges used, and the size of the overall stitch plan. 

\subsubsection{Models}
To create a fair benchmark for our technique, we replicate models found in four related papers, which also develop 3D structures from flat sheets~\cite{curveups-ruslan-guseinov-TOG17, narumi2023inkjet, inverse-design-inflatable-julian-panetta-TOG21,shrink-and-morph-David-Jourdan-TOG23}. We furthermore add six that would not be supported by those techniques as they produce 2.5D textures on materials or construct a more advanced topology from separate sheets stitched together. 

\subsubsection{Pre-processing}
OriStitch has a minimum resolution of each face edge. To accommodate for this, we remeshed each model to regular triangles, for very small models, we scaled the model up to overcome these limitations. 

14/28 models required some form of re-meshing, either because they originally were quad meshes or the faces were smaller than our minimum resolution. For 6/28 models the original scale was not evident from the source, in these cases we set the scale to be commensurate for our resolution.

In the 8/28 models that are not a topological disc, we place one cut (two in the case of the torus) to reduce the topological complexity and later sow these edges together. The flat fabrication process requires the input geometry to be a topological disc.

\subsubsection{Results}

\begin{figure*}[h]
  \centering
  \includegraphics[width=\textwidth]{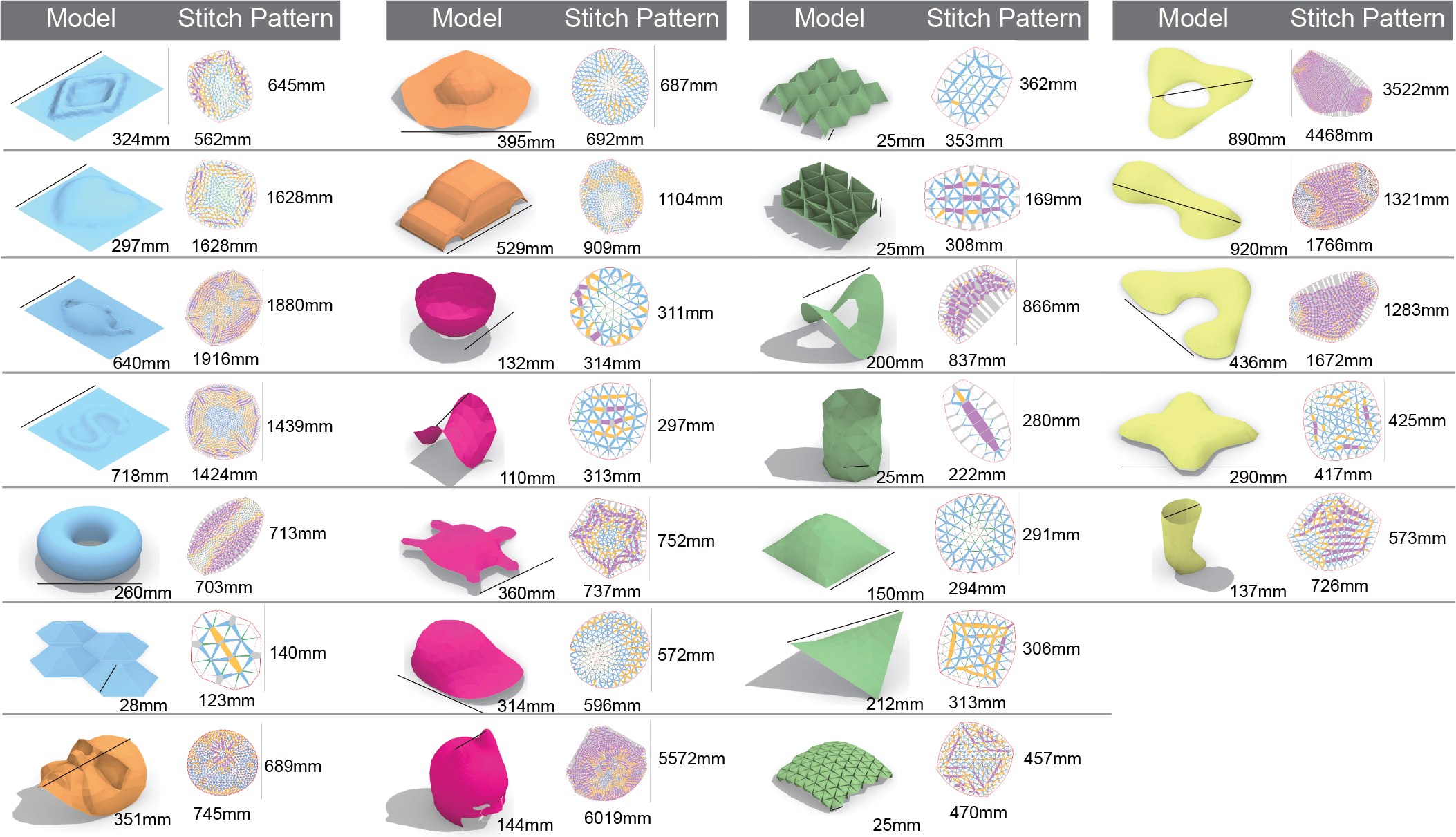}
  \caption{The results of our evaluation. The colors refer to the different origins of the papers: orange~\cite{shrink-and-morph-David-Jourdan-TOG23}, pink~\cite{curveups-ruslan-guseinov-TOG17}, yellow~\cite{inverse-design-inflatable-julian-panetta-TOG21}, green~\cite{narumi2023inkjet}, the blue models are added to show 2.5D textures and other topology (cut and restitched) converted by OriStitch. The color variation within the stitch patterns represent the different hinge types (type 0= green, type 1=blue, type 2 =orange and type 3=purple}
  \label{fig:results}
\end{figure*}

Figure~\ref{fig:results} shows the models and resulting stitch patterns generated by OriStitch. 26/28 of the models will correctly fabricate. The 2 models that did not work failed because the minimum resolution of Oristitch triangles and required re-triangulation at that scale degraded the original models to a degree that they no longer resemble the originals. 

Table~\ref{tab:results} shows the variety of types of hinges used by OriStitch to make these models fabricate (type 0,1,2,3) and the minimal edge length of the faces. The models which required scaling during pre-processing all have in common that the minimum edge length matches the resolution of OriStitch. 

\begin{table}
  \caption{hinge types and faces of each of the models used for technical evaluation.}
  \label{tab:commands}
  \begin{tabular}{llll} 
    \toprule
    model & faces & hinge types \\
    \midrule
    \texttt{Grid}& 443  & 239,232,126,51 \\
    \texttt{Heart}& 590  & 420,236,161,44 \\
        \texttt{Teapot}& 2455  & 912,1363,945,400 \\    
        \texttt{Letter}& 1880  & 766,1304,524,170 \\    
        \texttt{Torus}& 857; 2 cuts  & 475,180,248,351 \\    
        \texttt{Tessellation}& 24  & 12,14,3,0 \\
        \texttt{Face}& 798  & 242,693,208,37 \\
        \texttt{Hat}& 479  & 253,388,54,0 \\
        \texttt{Car}& 941  & 509,636,221,15 \\
        \texttt{Half Sphere}& 100  & 57,66,15,4 \\
        \texttt{Hyperboloid}& 82  & 12,81,13,4 \\
        \texttt{Turtle}& 312  & 88,172,94,85 \\
        \texttt{Bump Cap}& 340 & 169,264,61,0 \\
        \texttt{Mask}& 1795  & 303,1170,518,664 \\
        \texttt{Miura Fold}& 72  & 27,68,1,0 \\
        \texttt{Waterbomb}& 60  & 10,50,6,12 \\
        \texttt{Bauhaus paraboloid}& 140; 1 cut  & 29,57,23,75 \\
        \texttt{Yoshimura Pattern}& 28  & 8,19,1,5 \\  
        \texttt{Mountain}& 112 & 74,82,0,0 \\ 
        \texttt{Saddle}& 98  & 41,58,32,3 \\ 
        \texttt{Ronresch}& 424  & 212,192,137,65 \\ 
        \texttt{Annulus}& 2542; 1 cut  & 557,820,580,1762 \\ 
        \texttt{Two Bumps}& 1435  & 753,688,307,355 \\ 
        \texttt{Cashew}& 658  & 173,288,158,327 \\     
        \texttt{Lilium}& 185  & 78,152,24,6 \\ 
        \texttt{Cast}& 250;1 cut  & 76,147,67,62 \\       
    \bottomrule
  \end{tabular}
  \label{tab:results}
\end{table}

\subsubsection{Discussion}
In total, OriStitch successfully converted 26/28 meshes into fabricatable stitch patterns.

We found that in cases where the mean curvature is very low (local maximum or minimum) but the Gaussian curvature is not (like the pink saddle shape or the orange hat), which implies this shape is locally not developable, Origamizer places faces very far away from one another depending on the steepness of the curvature. This results in a cycle of type~3 hinges in OriStitch. Technically this can be fabricated, but because there is a lot of crimping that has to happen in a small area, it may produce an excess of tucked away material and thus a reduction of hinge tightness.

\subsection{OriStitch hinges function in a broad range of materials.} 

One of the core advantages of embroidery is that the additive textile manufacturing method allows it to work on a range of off-the-shelf textiles. We evaluated whether and how OriStitch's hinges are influenced by material variations. We applied a simple stitch pattern to each sample.

\subsubsection{Stitch pattern}
\begin{figure}[h]
  \centering
  \includegraphics[width=\linewidth]{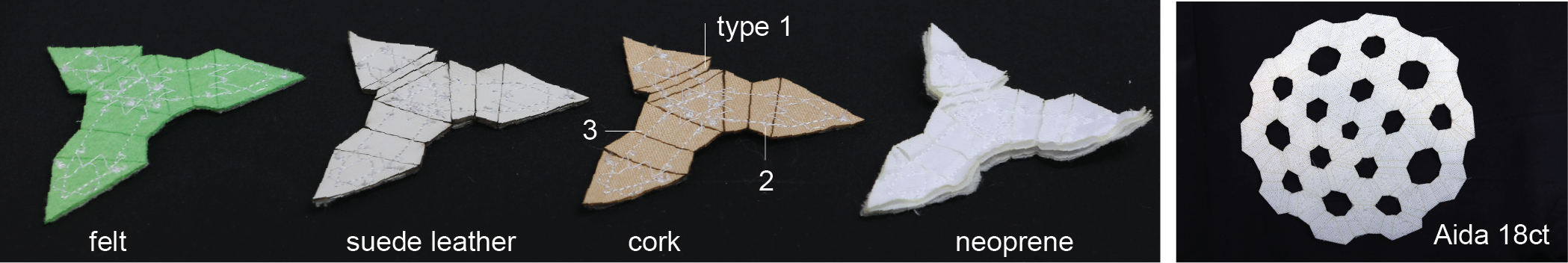}
  \caption{\textcolor{black}{The stitch pattern to evaluate the different materials. The different hinge types are labeled in the figure.}}
  \label{fig:testpattern}
\end{figure}

We used a basic stitch pattern as shown in Figure~\ref{fig:testpattern}, consisting of four fully connected faces (one being at the minimum resolution) and with three different hinge types to validate the full functionality of OriStitch hinges across materials. We also validate our hinges designs with a larger structure, here a hemisphere, consisting of the combination of several hinges. 

\subsubsection{Materials}
We selected the following materials: neoprene\footnote{https://burlapfabric.com/index.php?main\_page=product\_info\&products\_id=2933}, non-woven (leather)\footnote{https://www.amazon.com/dp/B08CF43C7X}, felt\footnote{https://www.amazon.com/dp/B01GCLS32M} , a composite material (natural cork with Polyurethane Mesh on the back)\footnote{https://www.amazon.com/dp/B0B4FNJPDW}, and even-woven fabric (Aida 18count)\footnote{https://www.michaels.com/product/loops-threads-aida-cloth-cross-stitch-fabric-295-x-36-18-count-10357725}.The thickness of the textile ranges from 0.3mm to 2mm. Table~\ref{tab:materials} shows an overview of the different samples.

\begin{table}[h]
    \centering
    \begin{tabular}{lll}
         \toprule
         material &  thickness &  composition \\
         \midrule
         suede leather & 1.4mm & natural  \\
         cork & 0.6mm & cork on Polyurethane  \\
         Neoprene & 2mm &  90\% Polyester \& 10\% Lycra Spandex \\
         felt& 1mm & acrylic  \\  
         Aida 18ct& 0.3mm & cotton  \\ 
    \end{tabular}
    \caption{materials used}
    \label{tab:materials}
\end{table}

\subsubsection{Results}
\begin{figure}[h]
  \centering
  \includegraphics[width=\linewidth]{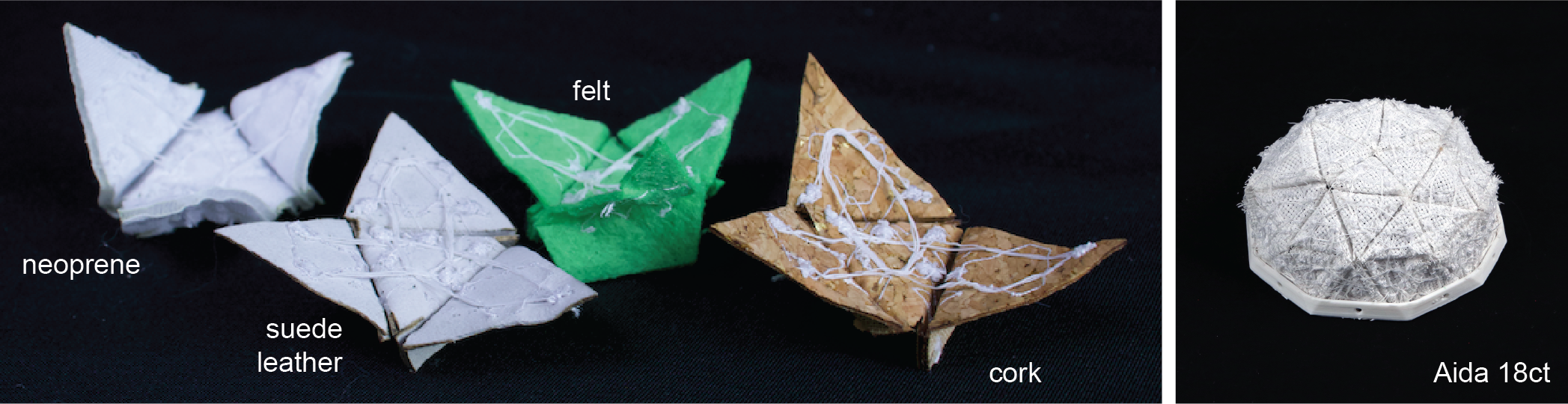}
  \caption{\textcolor{black}{Evaluated hinges and materials.}}
  \label{fig:material-tests}
\end{figure}

Figure~\ref{fig:material-tests} shows the results of our material testing. All hinges were fabricated in the different materials and managed to close. There was a slight gap in the composite cork material, likely because its larger thickness. For the Neoprene sample the hinges worked fine, but the tension of the shrinking threads on top of the face of the material caused some inward buckling of the face.

In fabrication, the Neoprene sample failed at first attempt because the widest hinge required too much travel between needle points pulling the material slightly during fabrication. Slowing down the machine for the second attempt produced a reliable result.

\subsubsection{Discussion}
OriStitch hinges worked reliably across material samples. For thicker materials, we do recommend using a thicker heat-actuated thread as the tension in the thread appeared to be the limiting factor for the thicker, less flexible materials. We furthermore saw that the material sample with a low density of fibers poses a risk of failure in the widest hinges. In our evaluation, we adjusted by reducing the speed of the machine, but we plan to incorporate a feature in the software to add additional needle points (reducing the bridging of the wide hinge) for more delicate materials.

Finally, when the materials are too soft, they are unable to withstand the tension within the faces. Our Neoprene still worked in the end, but some minimal surface stiffness appears to be required for predictable folding.

%% file: src/05_application.tex
\section{{Applications}}
\begin{figure}[h]
  \centering
  \includegraphics[width=\linewidth]{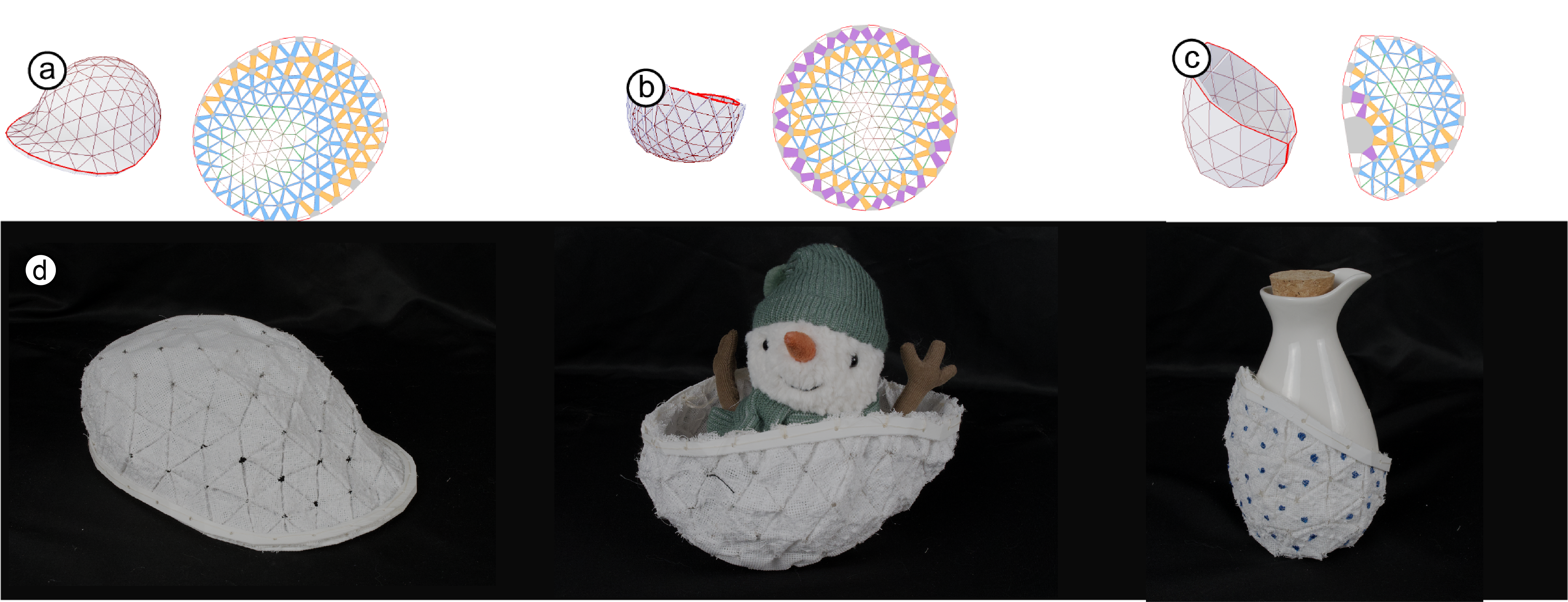}
  \caption{Conversion of application models: (a) cap, (b) bag, (c) vase cover, and (d) the final fabricated artifacts. }
  \label{fig:applications}
\end{figure}

We demonstrate OriStitch through a series of applications. In fabricating these, we note that the boundary edge of the model is not as tightly folded as the other hinges, this is a common problem for folded geometry as identified by \citet{inverse-design-inflatable-julian-panetta-TOG21}, to this end we sew a 3D printed rim of the geometry along the boundary of the mesh to provide outer faces with the required supporting force. Figure~\ref{fig:applications} shows the different application model, (a-c) their conversion from 3D to stitch pattern and (d) the final fabricated artifacts.

\textbf{Cap: custom fashion.} Fabric pleats have long been used in fashion items like clothes. Inspired by the tradition, we designed a cap. The design consists of 211 faces and 303 hinges. It was fabricated with 11 stages of hooping. This is the cap featured in Figure one. 

\textbf{Vase Cover: shape-matching existing geometry.} Textiles are often used in interior decoration, creating visual and textual effects as well as other functions like protection. The textile cover is designed to fit the organic ceramic vase, making it visually appealing while providing protection. The vase cover consists of 100 faces and 140 hinges, fabricated with 5 hoops. This model further demonstrates another advantage of the additive nature of Oristitch: that is combines excellently with other textile fabrication. Here the dots have been embroidered on the material beforehand. 

\textbf{Hand Bag: enclosing volumes.} The choice of fashion accessories reflects personal style and preferences, enabling individuals to express their unique tastes and create a distinctive look. This custom-shaped bag features 232 faces and 338 hinges, fabricated using 12 hoops. 

%\textbf{Wrist Rest: leather.}This wrist rest is made out of leather to be comfortable to the touch, yet with a smooth surface. The model consists of xxx faces, 66 hinges, and 3 hoops were used to fabricate this using Oristitch. 

%\subsection{Sound proof wall: doubly curved surfaces}
%This doubly-curved sound proof wall damps noise sound from getting to a table microphone. The natural textile properties of the felt used to fabricate this damp the sound well. The model consisted of xxx faces, 136 hinges, and was fabricated using 6 hoops.

%% file: src/06_discussion.tex
\section{Discussion}

Through the effort of designing OriStitch hinges, implementing computational support tools, evaluating the conversion tool, and demonstrating application examples, we gained valuable insights in the fabrication workflow. Some findings are outlined below.

\subsection{Forces at the Boundary} 
Our conversion tool assumes the faces of the model are rigid; this is a fair assumption as there are many stitches and threads spanning each face, making them relatively stiff. However, the faces and edges at the boundary of the sheet are still as flexible as the underlying fabric; we therefore reinforce the application models with a 3D printed rim. It would be interesting to add an additional stitch along the boundary to reinforce, or even stitch a channel to insert a non-fabric material (metal or plastic filament) in the rim after actuation. This will support the other (non-boundary) faces as well.

\subsection{Active thread}
For the purpose of accessibility of our approach, we went with an active thread that is available on the shelf and shrinks by 30\% and generates limited forces. Developing shrinking threads with higher shrinkage and stronger force would simplify the hinge designs, speed up fabrication process, and make it applicable to heavier materials. Additionally, the thread we are using is one-time shrinking; using revertible shrinking threads would open up additional possibilities. Researchers have explored the fabrication of custom fibers \cite{forman_fiberobo_2023}, but this exceeds the scope of this paper.

\subsection{Actuation Method}
In finding the ideal actuation method, we have tried many options, ranging from hair straighteners to ovens to oil baths. We generally found that OriStitch hinges are best actuated in boiling water simultaneous with heat gun post processing in sequence. The method ensures effective actuation with minimal manual effort. It is a typical problem in Origami that while the crease pattern will uniquely fold into the target shape, it is not always known that valid folding motions exist, let alone that they can all be folded simultaneously. Interesting follow-up work could investigate this path and/or explore other actuation methods like a heat gun on a robotic arm similar to the set up in \textit{FormFab}~\cite{mueller_formfab_2019}.

\subsection{Need of Guidance for Multi-hooping}
While OriStitch automatically generates the stitch and cutting files, there is still manual labor operating embroidery machines. Unless users have an industrial-type of machine, they will likely revert to multi-hooping, which requires elaborate alignment effort and frequent rewiring of threads. Right now, fabricating an OriStitch pattern requires intimate knowledge of the machine and materials. More work should be done to reduce this effort for machine-embroidery in general; most steps are repetitive, but easy to mix up, and it requires experience to oversee the machine to see when something is off. Better computer vision tools, as well as process guidance, could go a long way to make this process more accessible. 

\subsection{Size of Stitch Patterns} 
Multi-hooping proved to be instrumental in making larger sheets because of the limitations of the size of embroidery machines. However, especially for models with a lot of detail, the stitch patterns get \textit{very} large. Some models from our technical evaluation exceed the size of the laser cutter as well. While it is possible to laser-cut and stitch multiple sheets together, we note that depending on the fabric used, there is a limit to how much each hinge can carry. A future version of OriStitch should use the material properties and hinge layout to simulate how large a structure it supports, and maybe even locally reinforce, similar to FastForce~\cite{abdullah_fastforce_2021} for lasercutting.

\subsection{Advancing Embroidery Machines}
In this paper, we aimed to fabricate all models using an accessible home embroidery machine from Brother. We opted for this to minimize dependency on some rare machine properties and thus reduce the replicability of our approach. Obviously, larger machines will reduce the required number of hoops and time for fabrication; furthermore, machines with multiple needles allow fabricating more layers at once, and machines equipped with computer vision techniques further simplify the registration process. We did experiment with more advanced machines to confirm those capabilities. An extension of machines that would further increase the capabilities of our method would be integration with a laser cutter. Many machines have the ability to be equipped with cutting needles, but the resolution and orientation of the blade dramatically reduce its value for precision embroidery. 

%% file: src/07_futurework.tex
%\section{Future Work}

%\subsection{Leverage Embroidery}

%\subsection{Complex Geometry}

%\subsection{Heating Condition}

% Each author must be defined separately for accurate metadata
% identification.  As an exception, multiple authors may share one
% affiliation. Authors' names should not be abbreviated; use full first
% names wherever possible. Include authors' e-mail addresses whenever
% possible.

% Grouping authors' names or e-mail addresses, or providing an ``e-mail
% alias,'' as shown below, is not acceptable:
% \begin{verbatim}
%   \author{Brooke Aster, David Mehldau}
%   \email{dave,judy,steve@university.edu}
%   \email{firstname.lastname@phillips.org}
% \end{verbatim}

% The \verb|authornote| and \verb|authornotemark| commands allow a note
% to apply to multiple authors --- for example, if the first two authors
% of an article contributed equally to the work.

% If your author list is lengthy, you must define a shortened version of
% the list of authors to be used in the page headers, to prevent
% overlapping text. The following command should be placed just after
% the last \verb|\author{}| definition:
% \begin{verbatim}
%   \renewcommand{\shortauthors}{McCartney, et al.}
% \end{verbatim}
% Omitting this command will force the use of a concatenated list of all
% of the authors' names, which may result in overlapping text in the
% page headers.

% The article template's documentation, available at
% \url{https://www.acm.org/publications/proceedings-template}, has a
% complete explanation of these commands and tips for their effective
% use.

% Note that authors' addresses are mandatory for journal articles.

%% file: src/08_conclusion.tex
\section{Conclusion}
We presented OriStitch, a machine-embroidery workflow and software tool to create 3D self-folding textiles. We introduce a specific hinge design, which self-folds, is flexible in width, and fabricates reliably in a range of different materials. We have shown how this method allows us to create the cap of Figure 1. We evaluated our conversion and found that 26/28 models found in related papers are supported by our process, and identified how to extend the tool to address issues in the remaining five models. 

The OriStitch workflow makes creating intricate 3D textiles more accessible. In the future, we hope to see more workflows to reduce the labor required for the fabrication of expressive textile objects. Such custom fabrication used to be prohibitively expensive and solely accessible in high-fashion. Accessible workflows encourage exploration and inclusive design for both high-end and hobbyist users. 